\documentclass[prl,twocolumn,showpacs,floatfix]{revtex4}
\usepackage{graphicx} % enhanced support for graphics
\usepackage{color} % use colored text in latex
\usepackage{amsmath} % Defines extra environments for multiline displayed equations, as well as a number of other enhancements for math
\usepackage[urlcolor=blue, hyperindex, colorlinks, bookmarks=true,linkcolor=black,citecolor=black]{hyperref} % provides LaTeX the ability to create hyperlinks within the document
\usepackage{amssymb}% {bm}% bold math
\newcommand{\bra}[1]{\langle{#1}|}
\newcommand{\ket}[1]{|{#1}\rangle}

\begin{document}

\title{Stationary and transient leakage current in the Pauli spin blockade}
\date{\today}
\author{F. Qassemi}
\affiliation{Institute for Quantum Computing and Department of Physics and Astronomy, University of Waterloo, Waterloo, Ontario N2L 3G1, Canada}
\author{W. A. Coish}
\affiliation{Institute for Quantum Computing and Department of Physics and Astronomy, University of Waterloo, Waterloo, Ontario N2L 3G1, Canada}
\author{F. K. Wilhelm}
\affiliation{Institute for Quantum Computing and Department of Physics and Astronomy, University of Waterloo, Waterloo, Ontario N2L 3G1, Canada}
\begin{abstract}
We study the effects of cotunneling and a non-uniform Zeeman splitting on the stationary and transient leakage current through a double quantum dot in the Pauli spin blockade regime.  We find that the stationary current due to cotunneling vanishes at low temperature and large applied magnetic field, allowing for the dynamical (rapid) preparation of a pure spin ground state, even at large voltage bias.  Additionally, we analyze current that flows between blocking events, characterized, in general, by a fractional effective charge $e^*$. This charge can be used as a sensitive probe of spin relaxation mechanisms and can be used to determine the visibility of Rabi oscillations.
\end{abstract}


\maketitle

Initialization and readout of well-defined quantum states are necessary for spin coherence measurements \cite{spin-coherence}, single-spin resonance \cite{koppens:2006a,nowack:2007a,pioro-ladriere:2008a}, and quantum information processing.  Single electron spins in quantum dots show promise for quantum information tasks \cite{loss:1998a} due to their long coherence times, but their quantum states can be difficult to initialize (relying on slow spin relaxation processes) and read out.  The Pauli spin blockade (PSB) \cite{ono:2002a} partially solves these problems, where current through a double quantum dot (DQD) is blocked conditional on the microscopic spin state of electrons.

The PSB is, however, imperfect; hyperfine interaction between electron and nuclear spins in III-V semiconductors lifts spin selection rules and can lead to a finite leakage current \cite{koppens:2005a,hyperfine-sb}.   Very recently, PSB has been observed in DQDs made from silicon \cite{liu:2008a,shaji:2008a} and carbon nanotubes \cite{churchill:2008a,churchill:2008b}, in which the majority isotope has no nuclear spin.  Even in these systems, the PSB can be lifted through spin exchange with the leads due to, e.g., cotunneling processes \cite{liu:2005a,weymann:2008a,vorontsov:2008a}.  Significantly, cotunneling events have been shown to be essential even in nuclear-spin-carrying quantum dots to describe nuclear-spin polarization in the PSB regime \cite{baugh:2007a, rudner:2007a}, and therefore should be taken into account.

Single-spin resonance measurements often rely on the transient current that flows before current is blocked as a probe of the electron spin state \cite{koppens:2006a, nowack:2007a, pioro-ladriere:2008a}. An anomalously large transient current has recently been found \cite{koppens:2006a}, characterized by an effective charge $e^*$ that passes through a DQD between blocking events. Without a complete understanding of this additional leakage, it may not be possible to determine the visibility of Rabi oscillations in these systems.

Here, we evaluate both the stationary and transient leakage current through a DQD, giving simple analytical expressions for the stationary current and the transient effective charge $e^*$.  We find that $e^*$ reaches universal fractional values and that a measurement of $e^*$ in general can be used to extract valuable information related to slow spin relaxation processes.

\begin{figure}
\centering
 \includegraphics[width=0.45\textwidth]{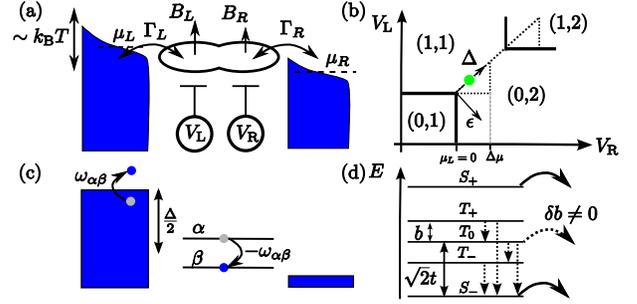}
\caption{\label{fig:doubledot}(Color online) A DQD coupled to leads (a).  The charge stability diagram is shown in (b).  At low temperature, inelastic cotunneling (c) induces transitions to lower-energy dot levels.  Two-electron states at $\epsilon=0$ are shown in (d) with allowed $T=0$ cotunneling transitions indicated with straight dashed arrows and curved arrows indicating sequential-tunneling processes.}	
\end{figure}
We consider a series-coupled DQD in a magnetic field gradient (Fig. \ref{fig:doubledot}(a)). A field gradient is important in spin resonance experiments for local addressing, and may arise from the stray field of a nanomagnet \cite{pioro-ladriere:2008a} or the Overhauser field due to non-uniformly polarized nuclear spins \cite{koppens:2006a}. We work in a regime where only the $(0,1)$, $(1,1)$, and $(0,2)$ charge states are relevant (the triangular region in Fig. \ref{fig:doubledot}(b)). Here, $(n_L,n_R)$ indicates $n_l$ electrons in dot orbitals $l=\{L,R\}$. The Hamiltonian in this projected subspace is $H=H_0+V$, where
\begin{eqnarray} 
H_{0} &=& \sum_j E_j\ket{j}\bra{j}+\sum_{lk\sigma}\epsilon_{lk}c_{lk\sigma}^\dagger c_{lk\sigma},\label{eq:H0Definition}\\
 V &=& \delta b\ket{S}\bra{T_0}+\sum_{kl\sigma j j'} t_{l}A_{l\sigma}^{jj'}c_{lk\sigma}\ket{j}\bra{j'}+\mathrm{h.c.}\label{eq:VDefinition}
\end{eqnarray}
The first term in $H_0$ describes the eigenstates $\ket{j}=\{\ket{\sigma},\ket{\alpha}\}$ of the unperturbed DQD, with single-electron states labeled by spin $\sigma=\{\uparrow,\downarrow\}$ and two-electron states labeled with $\alpha$, shown in Fig. \ref{fig:doubledot}(d) (three spin triplets, ($\alpha=T_{m_s}$, $m_s=0,\pm$) and two spin singlets, ($\alpha=S_\pm$), giving hybridized $(1,1)$ and $(0,2)$ charge states due to an interdot tunnel coupling $t$ \cite{states}).  The second term in $H_0$ gives the energy of Fermi-liquid leads. The Zeeman gradient $\delta b=g\mu_B (B_L-B_R)/2$  couples the $(1,1)$-singlet $\ket{S}$ and $m_s=0$ triplet $\ket{T_0}$ (here, $g$ is the $g$-factor and $B_l$ is a local magnetic field in dot $l$).  The second term in $V$ describes hopping processes from dot $l$ to lead $l$ with coupling $t_l$ and matrix elements $A_{l\sigma}^{jj'}=\bra{j}d^\dagger_{l\sigma}\ket{j'}$.  Here, 
$d_{l\sigma}^\dagger$ creates an electron in dot $l$ with spin $\sigma$ and $c_{lk\sigma}^\dagger$ creates an electron in lead $l$ and orbital $k$, with spin $\sigma$. 

Working from a standard Hubbard model for the DQD, we find the energies  $E_{\uparrow(\downarrow)}=-(\epsilon+\Delta)/2+(-)b/2$, where $\epsilon$ is the detuning (energy difference) between the $(1,1)$ and $(0,2)$ charge states and $\Delta$ controls the depth of the two-electron levels (see Fig. \ref{fig:doubledot}(c)).  For the two-electron states, we have $E_{T_0}=-\Delta$, $E_{T_\pm}=-\Delta\pm b$, and $E_{S_\pm}=-\Delta-\epsilon/2\pm\sqrt{\epsilon^2+8t^2}/2$, with $b=g\mu_B (B_L+B_R)/2$.

\begin{figure}
\centering
 \includegraphics[width=0.45\textwidth]{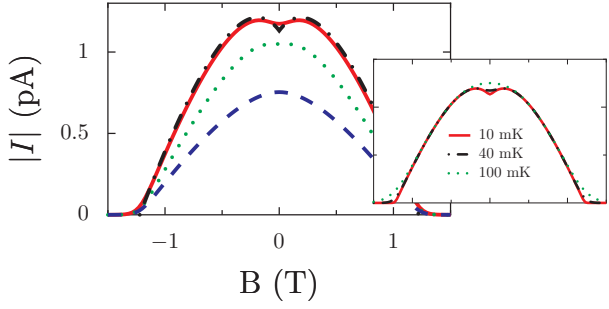}
\caption{\label{fig:current}(Color online) Leakage current in the PSB regime due to inelastic cotunneling processes.  We have taken $\epsilon=0$, $\Gamma_L=\Gamma_R=10\,\mu eV$, $t=100\,\mu eV$, $\Delta=1\,meV$, and $g=2.0$.  We show the evolution of $I(B)$ as $\delta B$ is varied from 0 mT (dashed line), to 20 mT (dotted line), 200 mT (solid line), and for $W_{T_0\to\sigma}\gg W_{\alpha\to\beta}$, $T=0$ (dash-dotted line, from Eq. (\ref{eq:IofBApprox})).  Evolution of the curve for $\delta B=200\,\mathrm{mT}$ as temperature is raised is shown in the inset.}
\end{figure}

We solve for the diagonal elements of the reduced (DQD) density matrix $\rho_{i}$ with the Pauli master equation:
\begin{equation}\label{eq:master-eq}
 \dot{\rho}_i = \sum_{j} (\rho_j W_{j\to i}-\rho_i W_{i\to j}).
\end{equation}
We calculate the transition rates $W_{i\to j}=W_{i\to j}^\mathrm{st}+W_{i\to j}^\mathrm{cot}$ directly from Fermi's golden rule. Here, the sequential-tunneling rates ($W_{i\to j}^\mathrm{st}=\sum_l W_{i\to j}^l \propto|t_l|^2$) describe direct hopping at leading order in the dot-lead coupling and the cotunneling rates ($W_{i\to j}^\mathrm{cot}\propto |t_l|^\eta,\,\eta>2$) are higher-order in $t_l$.

We consider standard initial conditions, with lead $l$ held in thermal equilibrium with Hamiltonian $H_0$ at chemical potential $\mu_l$. At first order in $V$, only the second term in Eq. (\ref{eq:VDefinition}) contributes to transport, giving the usual sequential-tunneling rates ($\hbar=1$) \cite{recher-golovach}:
\begin{eqnarray}
W_{\sigma\to\alpha}^l & = & \sum_{\sigma'}\Gamma_l |A^{\alpha\sigma}_{l\sigma'}|^2 f_l(\omega_{\alpha\sigma}),\nonumber \\
W_{\alpha\to\sigma}^l & = & \sum_{\sigma'}\Gamma_l |A^{\alpha\sigma}_{l\sigma'}|^2\left[1-f_l(\omega_{\alpha\sigma})\right].\label{eq:seq-rates}
\end{eqnarray}
Here, $\Gamma_l=2\pi\nu|t_l|^2$, where $\nu=\sum_k \delta(\epsilon_F-\epsilon_{lk})$ is the density of states per spin at the Fermi level, $f_l(E)$ is a Fermi function at temperature $T$ and chemical potential $\mu_l$, and $\omega_{ij}=E_i-E_j$.

For large bias, $\mu_L-\mu_R=\Delta\mu>|\Delta|>2k_B T$,  the stationary current is given by $I= e\sum_\alpha W_{\alpha\to\sigma}^R\bar{\rho}_\alpha$, where $\bar{\rho}_i$ is a solution to Eq. (\ref{eq:master-eq}) with $\dot{\rho}_i=0$. At leading order in $V$, current will be blocked if one of the triplet states is populated, since $W^R_{T_{m_s}\to\sigma}=0$.  This is the PSB effect.  In the absence of other spin-relaxation mechanisms, higher-order contributions in $V$ must be considered to explain a finite leakage current.  At second order in $V$ we find:
\begin{eqnarray}
 W_{T_0\to\sigma}^{R} &=& \Gamma_R\frac{\delta b^2}{8 t^2}\left[1-f_R(\omega_{T_0\sigma})\right],\label{eq:st-prime}\label{eq:field-assisted-st}\\
 W_{\alpha\to\beta}^{\mathrm{cot}} &=& \frac{2\Gamma_L^2}{\pi(\Delta-\epsilon)^2}\sum_{\sigma\sigma'\sigma''} |A_{L\sigma}^{\alpha \sigma''}|^2|A_{L\sigma'}^{\beta \sigma''}|^2 F(\omega_{\alpha\beta},T),\label{eq:cotunneling}
\end{eqnarray}
where $F(\omega,T)=\omega/(1-e^{-\omega/k_B T})$.  Eq. (\ref{eq:field-assisted-st}) gives the rate for field-assisted sequential-tunneling processes, where $\delta b$ converts $T_0$ to a singlet, which can then escape from the DQD via first-order (sequential) tunneling to the right lead (see the curved dashed arrow in Fig. \ref{fig:doubledot}(d)).  Eq. (\ref{eq:cotunneling}) gives the rate for an inelastic cotunneling process (Fig. \ref{fig:doubledot}(c)), allowing for conversion of triplets to singlets (dashed straight arrows in Fig. \ref{fig:doubledot}(d)). A competition between the two rates in Eqs. (\ref{eq:field-assisted-st}) and (\ref{eq:cotunneling}) will determine the leakage current in the PSB regime when other spin relaxation mechanisms due, e.g., to hyperfine and spin-orbit interactions are suppressed.  We note that Eq. (\ref{eq:field-assisted-st}) is independent of $\Gamma_L$, whereas Eq. (\ref{eq:cotunneling}) is independent of $\Gamma_R$, so an asymmetric coupling of the DQD to the leads will play a role in determining the relative scales of the two contributions.

  At high temperature ($k_B T>|\omega|$) we have $F(\omega,T)\simeq k_B T$, a regime that has been explored previously \cite{liu:2005a,vorontsov:2008a}. In this work, we focus on the low-temperature regime ($k_B T<|\omega|$), where $F(\omega,T)\simeq\omega\theta(\omega)$, giving rates that vanish linearly for small energy separation, with significant consequences (allowing, e.g., for the initialization of a pure spin state -- see below).  In Eq. (\ref{eq:cotunneling}), we have assumed $|\Delta-\epsilon|\gg|\omega_{\alpha\beta}|$ and have neglected resonant cotunneling contributions \cite{konig:1997a}, which are exponentially suppressed for $\Delta/2>k_B T$. Corrections due to spin exchange with the right lead are smaller in $\Delta/U'\ll 1$, where $U'$ is the interdot charging energy.  Additionally, we have considered the resonant tunneling regime \cite{vandervaart:1995a} ($\epsilon\lesssim t$).

We have numerically solved for the stationary current using the rates given in Eqs. (\ref{eq:seq-rates})-(\ref{eq:cotunneling}) and have plotted the result vs. $B=b/g\mu_B$ in Fig. \ref{fig:current}.  There is a sharp cutoff in the leakage current at large $b$ ($|b|>|E_{S_-}|$), which can be understood directly from Fig. \ref{fig:doubledot}(d).  When the lowest-energy triplet state ($T_-$ for $b>0$) is below the lowest-energy singlet ($S_-$), current will be blocked as soon as $T_-$ is populated, since the transition from $T_-$ to $S_-$ vanishes as $W_{T_-\to S_-}\propto\omega_{T_-,S_-}\theta(\omega_{T_-,S_-})$.  Thus, at low temperature a \emph{pure} spin state can be prepared ($\ket{T_+}$ or $\ket{T_-}$ depending on the sign of $b$).  We note that this preparation can be achieved even in the presence of a large bias $\Delta\mu> k_B T$.
This is a nontrivial result, since a large bias will generally drive the DQD out of equilibrium, resulting in a stationary state that is \emph{not} determined by thermal equilibrium with the leads \cite{excitation}.  Moreover, using this method a pure spin state can be dynamically prepared on a time scale $\tau_\mathrm{prep}\sim t^{-1} \left(\Delta/\Gamma_L\right)^2\sim 0.1\,\mu\mathrm{s}$ (using parameter values from the caption of Fig. \ref{fig:current}) without the need to wait for slow spin relaxation processes.

In Fig. \ref{fig:current}, we show cuts at $\epsilon=0$ describing the evolution of $I(B)$ as the field gradient $\delta B=\delta b/g\mu_B$ is increased from zero (see the auxiliary material \cite{supmat} for the dependence on $\epsilon$).  For sufficiently large $\delta B$, a dip appears near $B=0$.  Similar zero-field dips have been seen experimentally in several DQD systems and have been attributed to effects due to hyperfine \cite{koppens:2005a} or spin-orbit \cite{pfund:2007a,churchill:2008a} coupling.  In the present context, this zero-field dip can be understood from Fig. \ref{fig:doubledot}(d), without additional spin relaxation mechanisms.  When $\delta b$ (or $\Gamma_R$) is large, $T_0$ has a fast direct escape path by virtue of Eq. (\ref{eq:st-prime}), so only the $T_+$ and $T_-$ states can block current. At $b=0$, all triplets are degenerate, resulting in a vanishing inelastic cotunneling rate at low temperature ($W_{T_\pm\to T_0}\simeq 0$); transport can only occur if $T_\pm$ escapes via $S_-$.  However, for a small nonvanishing Zeeman splitting $b>0$, we have $W_{T_+\to T_0}\propto b\ne0$, allowing an additional escape route for $T_+$.  This results in an initial rise in current for small $b$, which eventually must fall to zero when $b\simeq\sqrt{2}t$, where $T_-$ goes below $S_-$.  In the limit where $W_{T_0\to\sigma}\gg W_{\alpha\to\beta}$, we find a simple expression for the stationary leakage current:
\begin{equation}\label{eq:IofBApprox}
 I = \frac{e}{\pi}\left(\frac{\Gamma_L}{\Delta}\right)^2\frac{(\sqrt{2} t-|b|)(\sqrt{2} t+3 |b|)}{\sqrt{2} t+|b|}\theta(\sqrt{2}t-|b|),
\end{equation}
where we have taken $\epsilon=0$, and $T=0$.  The current reaches a maximum at $b_\mathrm{max}=\pm \sqrt{2}t(2/\sqrt{3}-1)\approx \pm 0.22 t$.  Eq. (\ref{eq:IofBApprox}) is shown as a dash-dotted line in Fig. \ref{fig:current}.  We note that the limit $W_{T_0\to\sigma}\gg W_{\alpha\to\beta}$ required for Eq. (\ref{eq:IofBApprox}) can also be achieved for much smaller $\delta b$ when $\Gamma_R\gg\Gamma_L$.    A sufficiently large electron temperature will wash out the zero-field dip, but provided $b_\mathrm{max}\gtrsim k_{B}T$, this feature will still be visible (see the inset of Fig. \ref{fig:current}).  Reaching this regime for $T\simeq 100\,\mathrm{mK}$ should be possible in nanowire \cite{pfund:2007a} or nanotube DQDs \cite{graeber:2006a} where $t\gtrsim100\,\mu eV$ is common.

\begin{figure}
\centering
 \includegraphics[width=0.45\textwidth]{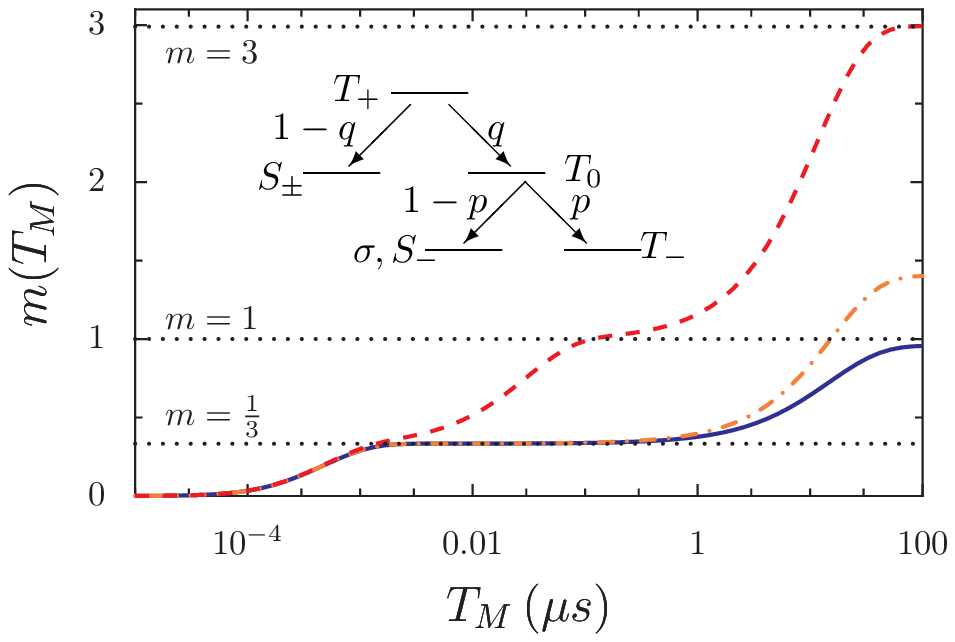}
\caption{\label{fig:mvalue}(Color online) Average number $m(T_M)$ of electrons passing through the DQD within measurement time $T_M$.  All values are as in Fig. \ref{fig:current} with the addition of $b=1.01\sqrt{2}t$, $T=0$, and $\Gamma_R=10\Gamma_L=10\,\mu eV$.  $m(T_M)$ is shown for $\delta B=0\,\mathrm{mT}$ (solid line), $\delta B=3\,\mathrm{mT}$ (dash-dotted line), and $\delta B=100\,\mathrm{mT}$ (dashed line). The predicted saturation points for 3 of 4 levels blocking ($m=1/3$), 2 of 4 levels blocking ($m=1$) and 1 of 4 levels blocking ($m=3$) are shown with dotted lines. The decay cascade (inset) defines the branching ratios $p$ and $q$.}
\end{figure}

We now turn to the transient (time-dependent) current that flows between blocking events.  We consider the instant after an electron has tunneled from the DQD to the right lead.  With spin-independent tunneling rates, this leaves the dot in an equal mixture of the states $\ket{\uparrow}$ and $\ket{\downarrow}$, setting the initial condition: $\rho_\sigma(0)=1/2,\;\rho_\alpha(0)=0$. The transient current into the right lead is then given by $I_R(t)=e\sum_\alpha W_{\alpha\to\sigma}^R \rho_\alpha(t)$.  The average number of electrons $m$ that passes through the DQD, given a charge collection (measurement) time $T_M$ is 
\begin{equation}
 m(T_M)=\frac{1}{e}\int_0^{T_M}d\tau I_R(\tau).
\end{equation}
In Fig. \ref{fig:mvalue} we plot $m(T_M)$ found by integration of Eq. (\ref{eq:master-eq}) for a range of parameters when the stationary current is zero (i.e., $k_B T=\epsilon=0$, $b>\sqrt{2}t$). The accumulated charge shows a series of plateaux at time scales determined by the three types of rates given in Eqs. (\ref{eq:seq-rates}), (\ref{eq:st-prime}), and (\ref{eq:cotunneling}).  To better understand these plateaux, we consider the long-time saturation value $m = \lim_{T_M\to\infty}m(T_M)$, which has been measured experimentally \cite{koppens:2006a} and can be evaluated directly.

We assume a probability $P_B$ for the DQD to be in a blocking state each time an additional electron tunnels onto the DQD.  The probability of exactly $n$ electrons passing through the DQD before current is blocked is then $P_n=(1-P_B)^n P_B$, from which we find $m=\sum_n nP_n=(1-P_B)/P_B$.  In the simplest case, there may be $N_B$  blocking levels out of $N$ total, giving $P_B=N_B/N$.  Assuming the two spin-polarized triplets $\ket{T_\pm}$ are perfect blocking states, we set $P_B=1/2$ since 2 out of 4 $(1,1)$-states block perfectly, giving $m=1$ (the expected value in Ref. \cite{koppens:2006a}; the measured value was $m\simeq 1.5$).  However, in the presence of some decay mechanism, transitions between the various two-electron states (see the inset of Fig. \ref{fig:mvalue}) no longer allow for a clear definition of ``blocking'' levels.  Nevertheless, we can still determine $P_B$ from the sum of probabilities for each independent path leading to a blocking state ($T_-$ for the case shown in Fig. \ref{fig:mvalue}): $P_B=P_{\sigma\to T_-}+P_{\sigma\to T_0\to T_-}+P_{\sigma\to T_+\to T_0\to T_-}=(1+p+qp)/4$, where $P_{A\to B\ldots}$ indicates the probability for a transition from state $A$ to $B$, etc., and where the branching ratios are given (for $b>0$) by: $p=W_{T_0\to T_-}/(W_{T_0 \to T_-}+W_{T_0 \to S_-}+\sum_{\sigma} W_{T_0 \to \sigma})$ and $q=W_{T_+\to T_0}/(W_{T_+\to T_0}+W_{T_+\to S_-}+W_{T_+\to S_+})$.  Inserting this result gives:
\begin{equation}\label{eq:m-pq}
 m = \frac{3-p-pq}{1+p+pq}.
\end{equation}
The average effective charge transported between blocking events $e^*=(m+1)e$ is non-integral in general, ranging from $e^*=\frac{4}{3}e$ to $e^*=4e$. Eq. (\ref{eq:m-pq}) allows for a precision measurement of slow spin-relaxation processes characterized by $p$ and $q$, \emph{independent} of the microscopic mechanism \cite{spinrelax}.  For concreteness, we consider the effects of cotunneling and field-assisted sequential tunneling below.

In the limit of zero detuning ($\epsilon=0$),  we find $q=1/2$, independent of $b$ and $t$, leaving
\begin{equation}\label{eq:minf-final}
 m = \frac{2-p}{\frac{2}{3}+p};\;\;p = \frac{2|b|}{\sqrt{2}t+2|b|+\frac{\pi}{2}\left(\frac{\Delta\delta b}{\Gamma_L t}\right)^2\Gamma_R}.
\end{equation}
Thus, at $\epsilon=0$, $m$ can be tuned from $m=3/5$ to $m=3$ by varying $b$, $\delta b$, $\Gamma_{L,R}$, and $t$.  Eq. (\ref{eq:minf-final}) correctly predicts the saturation values at $m$=0.96, 1.4, and 3.0 for the solid, dash-dotted, and dashed lines, respectively, in Fig. \ref{fig:mvalue}.

We have analyzed the effects of inelastic cotunneling and a magnetic field gradient on the PSB.  We find and explain a zero-field dip in the stationary current, which may help to explain recent experimental results \cite{koppens:2005a,pfund:2007a,churchill:2008a}.  We have shown that a pure spin state can be dynamically initialized, even at large bias, which is an important step on the way to full control over the quantum states of electron spins.  We have offered a possible explanation for an anomalously large value of the effective charge passing through the DQD found in experiments \cite{koppens:2006a}, which is important for single-spin resonance studies.  Our expression for this effective charge can be used to probe slow spin relaxation processes in the DQD to help understand the underlying physical mechanisms.  A fractional effective charge $e^*$ in transport is often taken as evidence of exotic electronic states \cite{frac-charge}. Here, we have shown that $e^*$ can reach universal fractional values in a simple system, without many-body correlations.

We thank D. G. Austing, J. Baugh, J. Gambetta, and D. Loss for useful discussions. We acknowledge funding from an NSERC discovery grant, QuantumWorks, IQC, WIN, CIFAR, and an Ontario PDF (WAC).

%%%%%%%%%%%%%%%%%%%%%%%%%%%%%% User specified LaTeX commands.
%Use the format (S#) for numbering equations in the supplementary material
\renewcommand{\theequation}{S\arabic{equation}}
\renewcommand{\thefigure}{S\arabic{figure}}
\renewcommand\bibnumfmt[1]{[S#1]}
\newcommand{\citenumfont}{S}
\def\thesection{S\Roman{section}}
\newpage
\cleardoublepage
\setcounter{figure}{0}
\setcounter{equation}{0}

\section{Auxiliary material for: ``Stationary and transient leakage current in the Pauli spin blockade''}

\subsection{Identifying spin decay}
In this supplement we show how to extract the magnetic-field dependence of microscopic relaxation processes at low field from the observed number of electrons that pass through the double dot between blocking events in the Pauli spin blockade regime. This measurement can be used to distinguish between spin-orbit, hyperfine-, and cotunneling-mediated spin relaxation mechanisms at low magnetic field, where other methods for single-spin detection fail.\cite{elzerman:2004a}

\begin{figure} [b]
	\centering
		\includegraphics[width=0.45\textwidth]{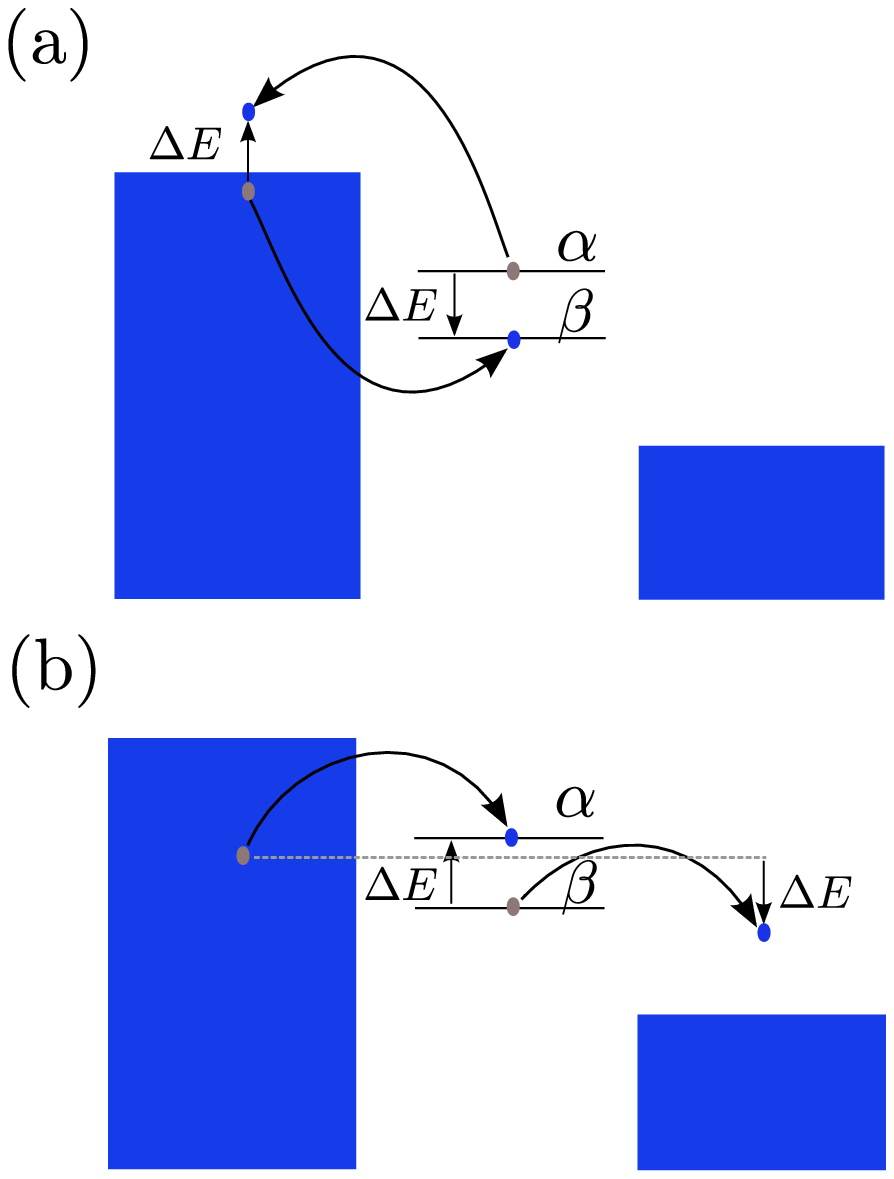}	
	\caption{\label{fig:processes} $\left|\alpha \right\rangle$ and $\left|\beta \right\rangle$ are double dot energy eigenstates and $\Delta E=E_{\alpha}-E_{\beta} > 0$. a) Available cotunneling processes b) Neglected cotunneling processes}
\end{figure}

We recall the definitions for the branching ratios $p$ and $q$ (given before Eq. (9) of the main text),
\begin{eqnarray}
p &=& \frac{W_{T_{0}\rightarrow T_{-}}}{W_{T_{0}\rightarrow T_{-}}+W_{T_{0}\rightarrow S_{-}}+\sum_{\sigma}W_{T_{0}\rightarrow \sigma}},\\
q &=& \frac{W_{T_{+}\rightarrow T_{0}}}{W_{T_{+}\rightarrow T_{0}}+W_{T_{+}\rightarrow S_{+}}+W_{T_{+}\rightarrow S_{-}}}.
\end{eqnarray}
These expressions can be simplified under certain experimental conditions.  In particular, we consider the case when there is no magnetic field gradient present, $\delta b=0$, and a sufficiently weak Zeeman splitting, so that transition rates from triplets to triplets (split by the Zeeman energy) are smaller than those from triplets to singlets (split by exchange) ($W_{T_{m}\rightarrow T_{m'}} \ll W_{T_{m} \rightarrow S_{\pm}}$). With $\delta b=0$, the field-assisted sequential-tunneling rates vanish:
\begin{equation}
W_{T_{0} \rightarrow \sigma}=0,
\end{equation}
and for a generic spin-flip Hamiltonian \cite{goldenrule},
\begin{equation}
W_{T_{+} \rightarrow T_{0}}=W_{T_{0} \rightarrow T_{-}}=W_{TT}.
\end{equation}

The rate $W_{T_{0} \rightarrow S_{-}}$ is independent of Zeeman splitting, $b=g\mu B$, since the $T_0-S_-$ splitting is independent of the global field.  In contrast, $W_{TT} \propto b^{\eta}$ for small magnetic field, depending on the spin-flip mechanism (as we have shown in the main text, $\eta=1$ for cotunneling-mediated spin-flips at low temperature, and previous work has shown $\eta=5$ for spin-orbit-mediated spin flips with phonon emission \cite{khaetskii:2001a}, and $\eta=3$ for hyperfine-mediated spin flips with phonon emission \cite{erlingsson:2002a}).  For sufficiently small Zeeman splitting $b$, the triplets become degenerate resulting in vanishing rates, allowing us to approximate
\begin{eqnarray}
\frac{W_{TT}}{W_{T_{+} \rightarrow S_{\pm}}} & , & \frac{W_{TT}}{ W_{T_{0} \rightarrow S_{-}}} \ll 1 \\
\Rightarrow p & \approx & \frac{W_{TT}}{W_{T_{0} \rightarrow S_{-}}} \ll 1\\
\Rightarrow q & \approx & \frac{W_{TT}}{\sum_{\alpha=\pm} W_{T_{+} \rightarrow S_{\alpha}}} \ll 1
\end{eqnarray}
In this regime, we approximate $m$ by its leading-order form in the small ratio $W_{TT}/W_{S_\pm}$:
\begin{eqnarray}
m &=& \frac{3-p-pq}{1+p+pq} = 3-2p+O(p^{2})+O(pq),\\
m &=& 3-2\frac{W_{TT}}{W_{T_{0} \rightarrow S_{-}}}+O\left(\left(\frac{W_{TT}}{W_{T \rightarrow S}}\right)^{2}\right).
\end{eqnarray}
Thus, by measuring $m(b)\cong 3-\gamma b^{\eta}$, it is possible to extract the relevant spin-flip mechanism: $\eta=$ 5,3, or 1 for spin-orbit interaction with phonon emission \cite{khaetskii:2001a}, hyperfine interaction with phonon emission \cite{erlingsson:2002a}, or cotunneling, respectively.

\section{Processes leading to dot excitation}
%\newpage{}
In our analysis we have neglected processes that can lead to excitation of the double dot at finite bias (see, e.g., Fig. \ref{fig:processes}(b) for an example).  These processes can lead to nonvanishing stationary populations of excited dot states, which would correspond to ``initialization errors'' in the scheme we have proposed.  However, we find that the leading excitation processes are suppressed by the small parameter 
\begin{eqnarray}
\frac{W^{b}_{\alpha \beta}}{W^{a}_{\alpha \beta}} \propto \left(\frac{\Delta}{U}\right)^2 \ll 1,
\end{eqnarray}
where $U$ is the energy cost for double occupancy of one of the dots.  Specifically, we have neglected virtual transitions to $(1,0)$-charge states. Here, $W^{a(b)}_{\alpha \beta}$ is the transition rate from $\left|\alpha \right\rangle$ to $\left|\beta \right\rangle$ due to process $a(b)$ in Fig \ref{fig:processes}. The purity of the initial state will be reduced by a correction of the order $\left( \frac{\Delta}{U} \right)^2$ which is negligible in our chosen regime.

\begin{figure}[t!]
\includegraphics[width=0.45\textwidth]{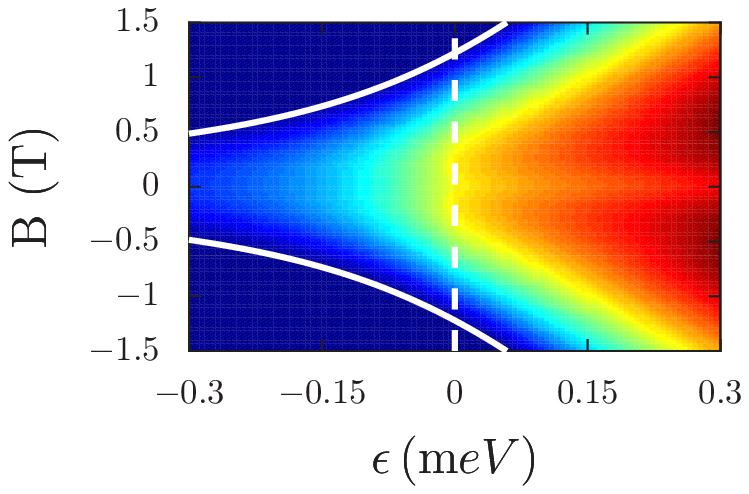}
\caption{\label{fig:bvsepsilon}Magnetic field and detuning dependence of 
leakage current in the spin-blockade regime. We have taken
$T = 40\,\mathrm{mK}$, $\mu_L = \mu_R = 10\,\mu eV$ , $t = 100\,\mu eV$ , 
$\Delta = 1\,\mathrm{m}eV$, $\delta B = 200\,\mathrm{mT}$, and $g = 2.0$. A 
density plot shows a suppression in the current at $B = 0$ and a sharp cutoff
when $|g\mu_B B| > |E_{S_-}|$ (solid white lines). The current runs
from $I = 0$ (dark blue) to $I = 1.7\,\mathrm{pA}$ (dark red).}
\end{figure}

\section{Detuning and field dependence of current}

In Fig. \ref{fig:bvsepsilon}, we show a density plot of the stationary current as a function of magnetic field and detuning for typical experimental parameters.  The current shows a suppression at $B=0$ corresponding to the zero-field dip in Fig. 2 of the main text.  Solid white lines are drawn to indicate when the lowest-energy triplet state becomes the ground state (i.e., when $|g\mu_B B| > |E_{S_-}|$). Fig. 2 of the main text corresponds to a cut along $\epsilon=0$ of Fig. \ref{fig:bvsepsilon}, indicated here with a white dashed line.


\begin{thebibliography}{28}
\bibitem{states}
\bibinfo{note}{Explicitly, the double-dot eigenstates are $\ket{T_+}=\ket{\uparrow\uparrow}$, $\ket{T_-}=\ket{\downarrow\downarrow}$, $\ket{T_0}=(\ket{\uparrow\downarrow}+\ket{\downarrow\uparrow})/\sqrt{2}$,  $\ket{S_{+}}=\cos\left(\frac{\theta}{2}\right)\ket{S}-\sin\left(\frac{\theta}{2}\right)\ket{S(0,2)}$, and  $\ket{S_{-}}=\sin\left(\frac{\theta}{2}\right)\ket{S}+\cos\left(\frac{\theta}{2}\right)\ket{S(0,2)}$, where $\ket{S}=(\ket{\uparrow\downarrow}-\ket{\downarrow\uparrow})/\sqrt{2}$, with $\ket{\sigma\sigma'}=d_{L\sigma}^\dagger d_{R\sigma'}^\dagger\ket{0}$ and $\ket{S(0,2)}=d_{R\uparrow}^\dagger d_{R\downarrow}^\dagger\ket{0}$. Here, $\tan\theta=-2\sqrt{2}t/\epsilon$.}
\bibitem{excitation}
\bibinfo{note}{We find that processes leading to DQD excitation in our considered large-bias regime are, however, suppressed by the small parameter $\Delta/U \ll 1$ \cite{supmat}, where $U$ is the single-dot charging energy.}
\bibitem{spinrelax}
\bibinfo{note}{For example, in the limit of small $b$, we find $m \simeq 3-\gamma b^\eta$, where $\eta =$ 5, 3, and 1 for spin-orbit, hyperfine-, and cotunneling-mediated spin relaxation, respectively \cite{supmat}.}
\end{thebibliography}
\begin{thebibliography}{4}
\bibitem{goldenrule}
\bibinfo{note}{We calculate golden-rule spin-flip rates assuming a perturbation of the form $V_{sf}= \sum_l S^{+}_l\Sigma_l +\Sigma^{\dagger}_lS^{-}_l$, where $S^\pm_l$ is a raising/lowering operator for the spin on dot $l$ and $\Sigma_l$ is a general (local) environment operator.}
\end{thebibliography}
\end{document}